\documentclass[12pt]{article}
\def\int {\intop\limits}
\newcommand{\be}{\begin{equation}}
\newcommand{\ee}{\end{equation}}

\newcommand{\aq}[1]{\label{#1}}
\newcommand{\barl}{\begin{array}{rl}}
\newcommand{\ear}{\end{array}}
\newcommand{\dst}[1]{\displaystyle{#1}}
\newcommand{\eq}[1]{eq.(\ref{#1})}
\newcommand{\Delbo}{\mbox{\boldmath${\Delta}$\unboldmath}}
\newcommand{\nubo}{\mbox{\boldmath${\nu}$\unboldmath}}

\newcommand{\rbo}{\mbox{\boldmath${\rho}$\unboldmath}}
\newcommand{\debol}{\mbox{\boldmath${\delta}$\unboldmath}}
\newcommand{\ebol}{\mbox{\boldmath${\eta}$\unboldmath}}
\newcommand{\sibol}{\mbox{\boldmath${\sigma}$\unboldmath}}
\def\fnote#1{\footnote}
\begin{document}
\vspace*{4.0cm}
\centerline{\Large {\bf Propagation of polarized high-energy }} 
\vskip .25cm 
\centerline{\Large {\bf photons in crystals }} 
\vskip .9cm 
\centerline{\large{\bf V. M. Strakhovenko}} 
\centerline{Budker Institute of Nuclear Physics, 
630090, Novosibirsk, Russia}
\centerline{e-mail address:v.m.strakhovenko@inp.nsk.su}
\vskip 1.5cm
\begin{abstract}
The density matrix is obtained for short-wavelength photons passing through
a medium. For sufficiently large thicknesses of a target this matrix is
expressed by the forward scattering amplitude of a photon. For 
multi-GeV photons scattering via virtual $e^{+}\,e^{-}$ pairs is the most 
important process. Its amplitude in a crystal is calculated. The results 
obtained are used to find the optimal thicknesses, orientations and types
of crystals for the circular~-~to~-~linear polarization 
conversion process.
\end{abstract}
\vskip 1.5cm
\section{ Introduction} 
High-energy circularly polarized photons, if available, would help in
solving several important problems in high energy physics like that of
spin-crisis (see e.g.\cite{Kirsebom} and literature cited there). If 
we start with linearly polarized multi-GeV photons, then they can be 
converted into circularly polarized ones provided that the suggestion 
of Cabibbo and collaborators \cite{Cabibo} to use a specially chosen 
crystalline plate for this purpose is true. This idea is now under 
experimental investigation at CERN within NA59 project, where the 
linearly polarized photons are also produced by means of a 
crystalline target from unpolarized electrons passing through it. 
From theoretical point of view, the conversion process itself is very 
interesting since for multi-GeV photons it is caused by the 
polarization of vacuum in the presence of the periodical electric 
field of a crystal.  This phenomenon has not been investigated so 
far.

To develop a description of the photon propagation, we start in Sec.2 from 
the Maxwell equations taking into account a current induced in a medium by the
incident wave in a rather general form. When the wavelength of a photon 
is much shorter than any other characteristic distance scale of the problem, 
the parameters of a wave packet change slowly while it propagates in a medium.
Using this fact,we
obtain the solution to the Maxwell equations in the short-wavelength
approximation, which is the analog of the eikonal approximation in the
fast particle scattering theory. The approximation used becomes valid starting
with a relatively small photon energy $\omega\,_{\sim}^{>}\,m$ , where $m$ is the
electron mass ( we use $\hbar\,=\,c\,=\,1$ units). Comparing our solution
with that of \cite{Newton} obtained for an amorphous medium, we find  that
a matrix, describing a change of the polarization and intensity of a wave
packet is deeply connected with the forward scattering amplitude. At least, it is
the case for large thicknesses, when this change becomes really prominent.
In the  multi-GeV energy region of interest, the photon 
scattering via virtual $e^{+}\,e^{-}$ pairs is the only process 
relevant to the problem. Its amplitude for a separate atom is well 
known ( see \cite{Preprint} for the forward scattering amplitude ). 
In crystals, along with this incoherent ( amorphous like ) 
contribution to the amplitude, the coherent one, caused by the 
periodicity of a lattice is present. The latter is calculated in Sec.3 by
means of the so called quasiclassical operator method. The details of this 
method along with many applications can be found in \cite{Book}. We use
 the results obtained to consider the
 circular~-~to~-~linear polarization
conversion process. Optimal ( according to the criterion formulated in 
\cite{Proposal} ) thicknesses and orientations are found for diamond, silicon and
germanium crystals for photon energy $\,\omega\,\sim\,100\,GeV$.

\section{ Propagation of short-wavelength photons}

 For the electric field ${\bf E}$ of a 
wave we obtain from the Maxwell equations
\be
\Big(\,\frac{\partial^{2}}{\partial t^{2}}\,-\,\nabla^{2}\Big)\,
E_{i}(x)=\,4\pi\,\int\,d^4x'\,R_{im}(x)\,\chi_{mj}(x,x')\,E_{j}(x')\, ,
\aq{Max} 
\ee 
where the operator $$\,R_{im}(x)\,=\,\frac{\partial^{2}}{\partial x_{i}
\partial x_{m}}\,-\,\delta_{im}\,\frac{\partial^{2}}{\partial t^{2}}\,.$$
When deriving \eq{Max}, the following relationship of the electric 
${\bf E}$ and the induced ${\bf D}$ fields was used:
\be
D_{i}(x)=\,E_{i}(x)\,+4\pi\,\int\,d^4x'\,\chi_{ij}(x,x')\,E_{j}(x')\, ,
\aq{DvE} 
\ee 
as well as the condition $\,\partial D_{i}(x)\,/\,\partial x_{i}\,=\,0\,$.
By definition, differentiating the integral in \eq{DvE} over $\,t\,$ ,
we obtain $\,{\bf j}(x)\,$ being a density of the induced current. When the
function $\,\chi_{ij}(x,x')\,$ depends  only on  the difference $\,t\,-\,t'\,$ ,
the equation for the vector potential $\,{\bf A}\,$ is essentially the same as
\eq{Max} for the electric field.

In the vacuum the solution to \eq{Max}, satisfying the condition
 $\,\partial E_{i}(x)\,/\,\partial x_{i}\,=\,0\,$, reads
\be
{\bf E}_{0}\,(x)\,=\,\int\,d{\bf k}\,g({\bf k}\,-\,{\bf k}_{0})\,{\bf e}
^{\perp}({\bf k})\,e^{-i kx}\, ,
\aq{E0} 
\ee
where $x$ and $k$ are 4-vectors: $x\,\equiv \,(t,{\bf r})$,
$k\,\equiv \,(\omega \,,\,{\bf k}\,)$ with $\,{\bf k}\,=\,\omega \nubo\,,
\nubo^{2}\,=\,1\,$, and 
${\bf e}^{\perp}({\bf k})\,$ being an arbitrary vector, perpendicular to ${\bf k}$. 
If we introduce $\delta_{ij}^{\perp}\,= \delta_{ij}\,-\nu_{i}\,\nu_{j}\,$, then
$\,e_{i}^{\perp}({\bf k})\,=\,\delta_{ij}^{\perp}\,e_{j}$. 
We assume that the function $\,g\,$ in
\eq{E0} vanishes except of the narrow region where $\,\mid{\bf k}\,-\,
{\bf k}_{0}\,\mid\,\ll\,\mid{\bf k}_{0},\mid\,$. So, the incident wave is
a wave packet propagating along $\,{\bf k}_{0}\,$. Let it encounter
a medium at some boundary . We try to satisfy  \eq{Max} in the medium 
using the field $\,{\bf E}\,$ in the  form of \be 
 E_{i}(x)\,=\,\int\,d{\bf k}\,g({\bf k}\,-\,{\bf k}_{0})\,F_{ij}({\bf r}\,,\,{\bf k})
 \,e_{j}^{\perp}({\bf k})\,e^{-i kx},
\aq{Ein} 
\ee
with $\,F_{ij}({\bf r}\,,\,{\bf k})\,=\,\delta_{ij}\,$ on the boundary. 
Assuming that $ \,\omega_{0} \,$ is sufficiently large, we expect that the function
$\,F_{ij}({\bf r}\,,\,{\bf k})\,$ varies very slowly with respect to $\,{\bf r}\,$.
Substituting  \eq{Ein} into \eq{Max} and neglecting the term $\,\nabla^{2}\,F\,$
( keeping only $ \,{\bf k}\partial / \partial {\bf r}\,F\,$), we obtain
\be
\barl
&\dst 
0\,=\,\int\,d{\bf k}\,g({\bf k}\,-\,{\bf k}_{0})\,\Big\{\,2i e^{-i kx}
k_{l}\frac{\partial F_{ij}({\bf r}\,,\,{\bf k})}{\partial r_{l}}\,+\\\\
&\dst 
+\,4\pi\,\int d^4x'\,R_{im}(x)\,\chi_{ml}(x,x')\,F_{lj}({\bf r'}\,,\,{\bf k})
 \,e^{-i kx'}\Big\}\,e_{j}^{\perp}({\bf k}).
\ear 
\aq{Feq} 
\ee
The integral $\,\int\,d^4x'\,$ in \eq{Feq} can be easily taken if we use the
Fourier-transforms:
\be
\barl
&\dst
\chi_{ml}(x,x')\,=\,\int\,\frac{d^4k_{1}d^4k_{2}}{(2\pi)^{8}}\,e^{i 
(k_{2}x'\,-\,k_{1}x)}\chi_{ml}(k_{1},k_{2})\,,\\\\
&\dst
F_{lj}({\bf r'}\,,\,{\bf k})\,=\,\int\,\frac{d{\bf q}}{(2\pi)^{3}}
e^{i {\bf q}{\bf r'}}F_{lj}({\bf q}\,,\,{\bf k})\, . 
\ear 
\aq{Fur} 
\ee
The most general form of the function $\,\chi_{ml}(k_{1},k_{2})\,$ in
crystals reads
\be
\chi_{ml}(k_{1},k_{2})\,=\,(2\pi)^{4}\,
\sum_{{\bf Q}}c_{ml}({\bf Q}\,,k_{1})\delta (k_{2}\,-\,k_{1}\,-\,Q),
\aq{Fcr} 
\ee
where $Q\,\equiv \,(0,{\bf Q})$ is a reciprocal lattice vector.
Using this representation, we obtain from  \eq{Feq}
\be
\barl
&\dst 
0\,=\,\int\,d{\bf k}\,g({\bf k}\,-\,{\bf k}_{0})\,e^{-i kx}\,\Big\{\,2i 
k_{l}\frac{\partial F_{ij}({\bf r}\,,\,{\bf k})}{\partial r_{l}}\,+\\\\
&\dst 
+\,4\pi\omega^{2}\delta_{im}^{\perp}\int\,\frac{d{\bf q}}{(2\pi)^{3}}
\sum_{{\bf Q}}c_{ml}({\bf Q}\,,k+q-Q)e^{-i {\bf Q}{\bf r}}\,F_{lj}
({\bf q}\,,\,{\bf k})\,e^{i {\bf q}{\bf r}}\Big\}\,e_{j}^{\perp}({\bf k}).
\ear 
\aq{Qfr} 
\ee
The second argument of the function $c_{ml}$ in \eq{Qfr} is
$k_{1}\,=\,k+q-Q\,\equiv \,(\omega,{\bf k+q-Q})$. This function 
 depends actually on $\omega_{1}\,=\,\omega \,$ and ${\bf 
n}_{1}\,=\,{\bf k}_{1}~/~\mid~{\bf k}_{1}~\mid 
\,\approx\,\nubo\,+\,{\bf s}^{\perp}$, where ${\bf s}\,=\,({\bf q-Q})/\omega$. We
have already used the fact that $\mid{\bf s}\mid\,\ll\,1$ neglecting it as compared
to unity.
In particularly, that is why we have $\,\delta_{im}^{\perp} \,$ in \eq{Qfr}
instead of $\,\delta_{im}\,-\,k_{1i} k_{1m}/\omega^2 \,$. We shall 
  see below that owing to this replacement the longitudinal 
  components of $\,F_{lj}({\bf r}\,,\,{\bf k})\,$ do not appear 
at any depth if 
they are absent on the boundary.
 However, when $\,\nubo\,$ is almost parallel to some crystal 
  axis, the transverse ( with respect to this axis ) component of 
  $\,\nubo\,$ should be compared with $\,\mid{\bf s}\mid\,$. We can 
  rule out $\,{\bf q-Q}\,$ from the second argument of the function 
  $c_{ml}$ when the angle of incidence $\,\vartheta _0\,$ with 
  respect to this axis ( transverse component of $\,\nubo\,$ ) is 
  sufficiently large:  $\,\vartheta _0\,\gg\,\mid{\bf s}\mid\,$.  In 
  what follows, we assume this condition to be fulfilled, bearing in 
mind that really it does not lead to serious limitations as a typical 
 magnitude of $\,\mid{\bf s}\mid\,$ at $\,\omega\,\simeq\,1\,GeV\,$ 
is $\,10^{-6}\,$.  The integral over $\,{\bf q}\,$ turns into 
$\,F_{lj}({\bf r}\,,\,{\bf k})\,$ according to the definition  
\eq{Fur} , when we change $\,c_{ml}({\bf 
Q}\,,k+q-Q)\,\Rightarrow\,c_{ml} ({\bf Q}\,,k)\,$ in \eq{Qfr}. This 
means  that the integral over $\,{\bf r'}\,$ in \eq{Feq} converges in 
the domain $\,\mid{\bf r}\,-\,{\bf r'}\mid\,\ll\,R\,$ where $\,R\,$ 
is a characteristic scale for the noticeable change of the function 
$F_{lj}({\bf r'})$. So that, we could substitute $F_{lj}({\bf 
r'})\,\Rightarrow\,F_{lj}({\bf r})$ yet in \eq{Feq}.  Now \eq{Qfr} 
goes over into 
\be 
\barl 
&\dst 
0\,=\,\int\,d{\bf k}\,g({\bf 
k}\,-\,{\bf k}_{0})\,e^{-i kx}\,\Big\{\,2i k_{l}\frac{\partial 
F_{ij}({\bf r}\,,\,{\bf k})}{\partial r_{l}}\,+\\\\ &\dst 
+\,4\pi\omega^{2}\delta_{im}^{\perp}
\sum_{{\bf Q}}c_{ml}({\bf Q}\,,k)e^{-i {\bf Q}{\bf r}}\,F_{lj}
({\bf r}\,,\,{\bf k})\,\Big\}\,e_{j}^{\perp}({\bf k}).
\ear 
\aq{Fkr} 
\ee
Here we are interested in the transverse ( with respect to $\,\nubo\,$ ) tensor
$\,F_{ij}\,$. Let us analyze, whether such a form  of the sought tensor is
consistent with \eq{Fkr} and boundary conditions.
According to \eq{Fkr}, the left longitudinal components of this tensor
$\, F_j^{\parallel}\,=\,\nu_i F_{ij}({\bf r}\,,\,{\bf k})\,$ are independent of
the penetration depth. If we suppose that $\, F_j^{\parallel}\,=\,0\,$ on the
boundary, then $\,F_{lj}\,\equiv\,\delta_{ln}F_{nj}\,=\,\delta_{ln}^{\perp}
F_{nj}\,$ at any depth. In particularly, this implies that only  transverse  
 components of the tensor $c_{ml}({\bf Q}\,,k)$ are present in 
\eq{Fkr}. The right longitudinal components of the sought tensor
do not enter \eq{Fkr} owing to the factor  $\,e_{j}^{\perp}({\bf k})\,$. So,
all the tensors below are self-consistently assumed to be transverse. They can
be presented as 
two-dimensional matrices. Remembering this, we obtain from \eq{Fkr}
\be
F({\bf r}\,,\,{\bf k})\,=\,\exp \Big\{\,2\pi i\omega z
\sum_{{\bf Q}}c({\bf Q}\,,k)e^{-i {\bf Q} \rbo}\int_{0}^{1}
\,dx\,e^{-i x z Q_{\parallel}}\Big\}\, ,
\aq{Frc} 
\ee
where $\rbo\,=\,{\bf r}\,-\,z\nubo\,$,$\,z\,=\,\nubo{\bf r}\,$,$\,
Q_{\parallel}\,=\,\nubo{\bf Q}\,$. For $Q_{\parallel}\,=\,0\,$ the 
integral in \eq{Frc} equals unity, while for 
$Q_{\parallel}\,\not=\,0\,$ it is of the order of 
$\,(Q_{\parallel}z)^{-1}\,$ . 
Noticeable effects appear when the main ( $Q_{\parallel}\,=\,0\,$ ) 
term in the phase of \eq{Frc} is of the order of unity. As shown in 
 the next Section, it happens for $z$ of several centimeters when
the contribution of terms with $\,Q_{\parallel}\,\not=\,0\,$ to this 
 phase $\,\sim\,(Q_{\parallel}z)^{-1}\,\ll\,1\,$ and can be  
neglected. Since matrices
$\,\sum_{{\bf Q}}c_{ml}({\bf Q}\,,k)e^{-i {\bf Q}{\bf r}}\,$   are in 
general non-commutative at different $z$ , the exact solution to \eq{Fkr}                           
is the   $z$ -ordered exponential function. However, corrections to
\eq{Frc} due to $z$ -ordering are connected with terms 
$\,Q_{\parallel}\,\not=\,0\,$ being as small ($\,(Q_{\parallel}z)^{-1}\,$ )
as terms already neglected.
  It must be remembered also that ( for the coherent yield )
$\,{\bf Q}\,$ is a discrete 
vector and that large values of $\,\mid{\bf Q}\mid\,$ do not 
contribute to the sum in \eq{Frc}. These means that for sufficiently 
thick crystals we should keep only terms with $Q_{\parallel}\,=\,0\,$ 
in the sum of \eq{Frc}:  \be F({\bf r}\,,\,{\bf k})\,=\,\exp 
\Big\{\,2\pi i\omega z \sum_{{\bf Q}_{\perp}}c({\bf 
Q}_{\perp}\,,k)e^{-i {\bf Q}_{\perp} \rbo}\Big\}\,.  \aq{Flb} \ee The 
condition $Q_{\parallel}\,=\,0\,$ implies that ${\bf 
Q}_{\perp}\,=\,0\,$ too, if $\,\nubo\,$ is not aligned on some 
crystal plane. More precisely, let $\,\psi\,$ be the angle of 
$\,\nubo\,$ with respect to the plane. For $\,\psi\,\gg\, 
\psi_{0}\,=\,d_{pl}/z\,$ , with $\,d_{pl}\,$ being the inter-planar distance,
we can retain the single term $\,c({\bf Q}\,=\,0\,,k)\,$ of the sum in \eq{Flb}. 
The angle $\,\psi_{0}\,$ is extremely small, even for very thin films. The 
integration
over $\,\nubo\,$ in \eq{Ein} should smear out a dependence of the general result
 on \rbo. We are not interested here whether such a dependence is observable
 or not. In what follows, the sum 
$\,\sum_{{\bf Q}_{\perp}}c({\bf Q}_{\perp}\,,k)
e^{-i {\bf Q}_{\perp} \rbo}\,$ in \eq{Flb} will be replaced by the matrix
$\,c({\bf Q}\,=\,0\,,k)\,\equiv\,c(0\,,k)\,$ which coincides with the average of 
this
sum  over $\,\rbo\,$. The matrix $\,c(0\,,k)\,$
has no singularities and is practically 
constant in the region $\,\psi\,<\,\psi_{0}\,$. So that, any directions  
are allowed for $\,\nubo\,$.
 
Keeping  in
the sum $\,\sum_{{\bf Q}_{\perp}}c({\bf Q}_{\perp}\,,k)
e^{-i {\bf Q}_{\perp} \rbo}\,$
 the only term with $\,{\bf Q}_{\perp}\,=0\,$ 
, we disregard
possible small-angle photon scattering. In this approximation there 
is no exchange between fractions of the wave-packet having different
$\,\nubo\,$,i.e., these  photons propagate
independently. So that,  for given direction, we can go over to the 
conventional idealization of a monochromatic plane wave. 
. Then the electric 
field is obtained from \eq{Flb} and \eq{Ein} with the change in the 
latter $\,g({\bf k}\,-\,{\bf k}_{0}) \Rightarrow\,\delta ({\bf 
k}\,-\,{\bf k}_{0})\,$ and then $\,{\bf k}_{0}\Rightarrow\,{\bf k}\,$ 
\be
E(x)\,=\,e^{-i \omega t}\,e^{\,i \omega z 
n}\,e^{\perp}({\bf k})\, ,
\aq{Epl} 
\ee
where the two-dimensional vectors $\,E(x)\,$ and $\,e^{\perp}({\bf k})\,$ are 
correspondingly the electric field in a medium and the polarization vector of the 
incident wave. The quantity $\,n\,= \,\delta_{ij}^{\perp}\,+2\pi c(0\,,k)_{ij}\,$ 
is a complex matrix representing, by definition, the index of refraction.
It depends on $\,\omega\,$ and ( in crystals ) on  
$\,\nubo\,$.  We introduce a matrix $\,\Pi\,= \,4\pi \omega^{2}c(0\,,k)\,$, then
$\,n\,= \,I\,+\Pi /(2 \omega^{2})\,$. The photon density matrix at depth $\,L\,$
reads
\be
\rho\,(L)\,=\,\exp \left\{iL\frac{\Pi}{2 \omega }\right\}\rho(0)
\exp \left\{-iL\frac{\Pi^{\dagger}}{2 \omega }\right\} \, ,
\aq{Rho} 
\ee
where $\,\rho(0)\,$ is the density matrix of the incident photon.
 
Comparing \eq{Epl} with the corresponding result of \cite {Newton}
obtained for an amorphous medium by the direct summation of fields coming from 
individual scatterers, we find that in this case $\,\Pi\,= \, 4\pi f(0)\,N\,$. 
In this formula $\, f(0)\,$ is the forward scattering amplitude for an individual 
particle ( the cross section $\,d\sigma/d\Omega\,=\,\mid f(\Delbo)\mid^2\,$ ,
\Delbo$\,= \,{\bf k}_{2}\,-\,{\bf k}_{1}\,$ )
and $\,N\,$ is a number of particles per unit volume. So, to find the
polarization operator $\,\Pi\,$, we have to calculate the forward scattering 
amplitude normalized in such a way, that it reproduces $\, 4\pi f(0)\,$ for an 
individual particle. Moreover, using the amplitude ( see Sec.3 for its 
normalization ) $\,T(k_{1},k_{2})\,$ for an arbitrary momentum transfer
\Delbo\quad, we obtain for $\,\chi(k_{1},k_{2})\,$
defined in \eq {Fur} 
\be
\chi(k_{1},k_{2})\,= \,T(k_{1},k_{2})/(2\omega^{2})\,.
\aq{Hit} 
\ee
From this relation and \eq{Fcr}, we can extract $\,c_{ml}({\bf Q}\,,k_{1})\,$
and then, if needed, take into account terms with $\,{\bf Q}\,\neq\,0\,$ omitted 
at the transition from  \eq{Frc} to \eq{Epl}.

As any other $\,2\times 2\,$ matrix,  $\,\Pi/(2 \omega )\,$ in \eq {Rho} can be
presented in the form of $\,\Pi/(2 \omega )\,\equiv \,a + {\bf b}\sibol\,$ ,
where \sibol \quad are the Pauli matrices. Note that the eigenvalues of the
matrix $\,a + {\bf b}\sibol\,$ are  $\,a\,\pm\,\sqrt {{\bf b}^2}\,$. Using a
similar representation for the
initial density matrix $\,\rho(0)\,=\,(I + \ebol\sibol\,)/2\,$ , with \ebol\quad
 being the
initial Stokes vector, we find from \eq {Rho}
\be
\barl
&\dst
\rho\,(L)\,=\,\frac{1}{2}e^{-\gamma L}\Big\{\,\frac{1+\ebol\sibol}{2}\left(
\cosh \alpha+\cos \beta \right)+\left(
\cosh \alpha-\cos \beta \right)
\Big[\frac{1-\ebol\sibol}{2}\left({\bf g}_1^2 +
{\bf g}_2^2 \right)\,+\\\\
&\dst 
+\left({\bf g}_1 \times {\bf g}_2\,,\,\sibol-\ebol \right)
+\left({\bf g}_1\ebol \right) \left({\bf g}_1 \sibol \right)+
\left({\bf g}_2\ebol\right)\left({\bf g}_2\sibol\right)\Big]
\,-\\\\
&\dst 
-\left[
\left({\bf g}_2\,,\,\sibol+\ebol\right)+\left({\bf g}_1\times\ebol\,,\,
\sibol\right)\right] \sin \beta -\left[
\left({\bf g}_1\,,\,\sibol+\ebol\right)-\left({\bf g}_2\times\ebol\,,\,
\sibol\right)\right] \sinh \alpha \Big\}\quad, 
\ear 
\aq{Rof} 
\ee
where $\,\gamma\,=\,2Im (a)\,$ , $\,\alpha\,=\,2 L\,Im (\sqrt {{\bf b}^2})\,$ ,
$\,\beta\,=\,2 L\,Re (\sqrt {{\bf b}^2})\,$ , $\,{\bf g}\,\equiv\,{\bf g}_1+
i{\bf g}_2\,=\,{\bf b}/ \sqrt {{\bf b}^2}\,$. The real vectors $\,{\bf g}_1\,$
 and $\,{\bf g}_2\,$ satisfy the conditions $\,{\bf g}_1{\bf g}_2\,=\,0\,$ ,
 $\,{\bf g}_1^2-{\bf g}_2^2\,=\,1\,$ , since $\,{\bf g}^2\,=\,1\,$. As
$\,tr\, \rho(0)\,=\,1\,$ , the fraction of outgoing photons is given by
$\,FRAC\,=\,tr\, \rho(L)\,$. The Stokes vector at the depth $\,L\,$ is
$\,\ebol(L)\,=\,tr\, \sibol\rho(L)/FRAC\,$. In the explicit form:
\be
\ebol(L)\,=\,\frac{{\bf S}}{P}\quad, \quad FRAC\,=\,e^{-\gamma 
L}\,P\quad,\quad \ebol^2(L)\,=\,1\,-\frac{1\,-\ebol^2}{P^2}\quad,
\aq{etaf} 
\ee
where
\be
\barl
&\dst
P\,=\,{\bf g}_1^2 \cosh \alpha\,-\,{\bf g}_2^2 \cos \beta\,-\,
\left({\bf g}_1\,,\,\ebol\right) \sinh \alpha \,-\,
\left({\bf g}_2\,,\,\ebol\right) \sin \beta \,-\\\\
&\dst 
( \cosh \alpha-\cos \beta )
({\bf g}_1 \times {\bf g}_2\,,\,\ebol )\,; \\\\
&\dst 
{\bf S}\,=\,\ebol\, \left({\bf g}_1^2 \cos \beta\,-\,
{\bf g}_2^2 \cosh \alpha \right)\,-\,
\left[{\bf g}_1\,-\,{\bf g}_2 \times \ebol \right] \sinh \alpha\,-\,
\left[{\bf g}_2\,+\,{\bf g}_1 \times \ebol \right]  \sin \beta\,+\\\\
&\dst
[{\bf g}_1 \times {\bf g}_2\,+\,{\bf g}_1 
\left({\bf g}_1\,,\,\ebol\right)\,+\,{\bf g}_2
\left({\bf g}_2\,,\,\ebol\right)](
\cosh \alpha-\cos \beta )\quad.
\ear 
\aq{expl} 
\ee
We emphasize that, according to \eq {etaf}, the polarization degree
$\,\mid \ebol(L) \mid\,$ is an increasing function of depth $\,L\,$
except for $\,\ebol^2\,=\,1\,$ or $\,P^2\,=\,1\,$. 
    
\section{ Scattering of photons in a crystal}

 We now assume that $\,\omega\,\gg\,m\,$. Generally
speaking, the amplitude of the forward Compton scattering off electrons
$\,f_{C}(0)\,$ may be of comparable size with that due to  virtual 
$e^{+}\,e^{-}$ pairs ( VP ) even in $\,GeV\,$ energy region ( see discussion 
in \cite{Preprint} ). However, we can neglect $\,f_{C}(0)\,$ in what follows since 
it is, first of all, the unit  matrix  multiplied by some scalar factor
and, second, real 
. That is why its 
contributions to $\, \Pi\,$ and $\, \Pi^{\dagger}\,$ in \eq{Rho} are cancelled. 
For the same reason we can omit the real part of the incoherent contribution to the
forward scattering
VP amplitude. Nevertheless, its imaginary part, which does not cancel in \eq{Rho},
will be taken into account. It is the unit  matrix multiplied by
$\,\omega\,W_{p}\,$, with $\,W_{p}\,$ being the probability  per unit length  
of the $e^{+}\,e^{-}$ pair production by a photon.
Remember that the incoherent probability $\,W_{p}\,$  in crystals is 
smaller than in corresponding amorphous media (see ,e.g. \cite 
 {Book}).  So, we must calculate here only the coherent contribution 
to the VP amplitude.

The properly normalized transition amplitude of the photon with 4-momentum 
$\,k_{1\mu}\, 
=\,(\omega_1,{\bf k}_1)\,$ and polarization vector $\,e_{1\mu}\,$ 
into the photon with $\,k_{2\mu}\,= (\omega_2,{\bf k}_2)\,$, 
$\,e_{2\mu}\,$ reads
\be 
T(k_1\,,\,k_2)\,=\,2i \alpha\,\int\,d^4x_1d^4x_2
\,Tr\left[\,G(x_2,x_1)\,\hat e_1\,e^{-ik_1x_1}\,G(x_1,x_2)\,\hat 
e_2^{\ast}\,e^{ik_2x_2}\right]\,, 
\aq{Tin} 
\ee 
where $\alpha\,=\,1/137\,$ is the fine structure constant.   The electron 
Green function $\,G(x_1,x_2)\,$ can be expressed via solutions 
$\,\Psi_n^{(\pm)}(x)\,$ to the Dirac equation in the corresponding 
external field  
\be 
iG(x_2,x_1)\,=\,\theta(t_2-t_1)\,\sum_n\,\Psi_n^{(+)}(x_2)\,
\bar{\Psi}_n^{(+)}(x_1)\,-\,\theta(t_1-t_2)\,\sum_n\,\Psi_n^{(-)}(x_2)
\,\bar{\Psi}_n^{(-)}(x_1)\,.
\aq{Green}
\ee
Recollect that ( see discussion in \cite {M2} ) for $\,\omega\,\gg\,m\,$    
the contribution to the amplitude~, in terms of the non-covariant perturbation 
theory, is 
given by the diagram where the pair production by the initial photon 
precedes the annihilation of this pair into the final photon . 
We find, keeping only terms proportional to $\,\theta(t_2-t_1)\,$
\be
\barl
&\dst 
T(k_1\,,\,k_2)\,=\,2i\alpha\,\sum_{n,m}\,\int\,\,d^4x_1d^4x_2\,\theta(t_2-t_1)\,
V_{nm}(x_1,e_1,k_1)\,V_{nm}^+(x_2,e_2,k_2)\,\,,\\\\
&\dst
V_{nm}(x,e,k)\,=\,\bar{\Psi}_n^
{(+)}(x)\,\hat{e}\,e^{-ikx}\,\Psi_m^{(-)}(x)\,\,.
\ear
\aq{Tkk}
\ee

It is really important that in the Fourier-transform of a crystal potential
$$U({\bf q})\,=\,\int\,d{\bf r}\,e^{-i{\bf q}{\bf r}}\,U({\bf r})$$  
only small momenta $\,\mid {\bf q}\mid \,\ll\,m\,$ are present. For such
potentials, the  quasiclassical operator method can be applied in its standard
form to calculate the amplitude $\,T(k_1\,,\,k_2)\,$. Then   
further transformations in \eq{Tkk}
are completely identical to those leading to the quasiclassical 
expression for the probability of $\,e^+e^-\,$ pair production by a 
photon in an external field ( see Sec.3 in \cite{Book}), because
 $\,V_{nm}\,$ in \eq{Tkk} is essentially the matrix element for this process.
Notice that the amplitude  $\,T(k_1\,,\,k_2)\,$ is  the contraction 
of $\,e_{1\mu}e_{2\nu}^{\ast}\,$ with the sought tensor 
$\,T^{\mu\nu}(k_1\,,\,k_2)\,$.  For the transverse ( with respect to 
$\,\nubo_1\,=\, {\bf k}_1/\omega_1\,$ ) components of this tensor we 
obtain from \eq{Tkk} \be \barl &\dst 
T^{ij}=\frac{i\alpha}{(2\pi)^2}\delta(\omega_2
-\omega_1)\int d{\bf r}\,exp(-i\Delbo {\bf r})
\int_{0}^{\infty}d\tau
\int_{0}^{\omega}d\varepsilon\Big[\frac{\omega}{\varepsilon(\omega-
\varepsilon)}\Big]^2\int d{\bf p}_{\perp}(0)e^{-iA}\,B^{ij},\\\\
&\dst
A\,=\,\frac{\omega m^2}{2\varepsilon(\omega-\varepsilon)}\,
\left[\,\tau\,+\,\int_{-\tau/2}^{\tau/2}\,\frac{ds}{m^2}
{\bf p}_{\perp}^2\,(s)\,\right]\,;\\\\
&\dst
B^{ij}\,=
\,\delta_{\perp}^{ij}\left[\,m^2\,+\,\left({\bf p}_{1\perp}{\bf p}_{2\perp}
\right)\right]\,+\,
\left(\frac{2\varepsilon\,-\omega}{\omega}\right)^2
p_{1\perp}^i\,p_{2\perp}^j\,-\,p_{2\perp}^i\,p_{1\perp}^j\,\,,
\ear
\aq{Tes}
\ee
where $\,{\bf p}_{1,2}\,\equiv\, {\bf p}(\mp\tau/2)\,$ being
the momentum of an electron on a classical trajectory at the corresponding time,  
$\,\Delbo\,=\,{\bf k}_2\,-\,{\bf k}_1\,$. When the external field vanishes,
the amplitude $\,T\,$ must vanish as well. It assumes  a subtraction which
really affects only the term proportional to $\,m^2\,$ in $\,B^{ij}\,$.
This subtraction will be performed in the explicit form below.

Generally speaking, for further calculations we have to know the 
dependence of $\,{\bf p}_{\perp}(t)\,$ on time ,i.e. the 
corresponding mechanical problem should be solved. For arbitrary 
potential it cannot be done analytically. Fortunately, it is 
sufficient to know the alteration of the quantity $\,{\bf 
p}_{\perp}(t)\,$ during the process formation time $\,\tau_f\,$. In 
\eq{Tes} it is the characteristic size of the variable $\,\tau\,$  
contributing to the integral. 
In the rectilinear trajectory approximation (RTA),  we 
substitute into equations of motion $\,{\bf r}\,+\,\nubo_1\,t\,$ instead
of the exact solution $\,{\bf r}(t)\,$. We refer for details to \cite {Book}
stating here that in crystals the amplitude $\,T\,$ can be calculated
by the use of RTA  for arbitrary photon energy ($\,\omega\,\gg\,m\,$) and
crystal orientation.

Representing the crystal potential $\,U({\bf r})\,$ as a sum over vectors
$\,{\bf q}\,$   of the reciprocal lattice  
$$U({\bf r})\,=\,\sum_{\bf q}G({\bf q})\,e^{i{\bf q}{\bf r}}
\,\,,$$ we find within RTA for the transverse momentum
\be 
{\bf p}_{\perp}(t)\,=\,{\bf p}_{\perp}(0)\,+\,\debol(t)
\,\equiv\,{\bf p}_{\perp}(0)\,-\,\sum_{\bf q}G({\bf q})\,e^
{i{\bf q}{\bf r}}\,\frac{{\bf 
q}_{\perp}}{q_{\parallel}}\,\left(\,e^{iq_{\parallel}t}\,-\,1\, 
\right)\,,
\aq{Del} 
\ee 
where $\,q_{\parallel}\,=\,(\nubo_1 {\bf q})\,$. 
Substituting \eq{Del} into \eq{Tes}  , we can take the 
Gaussian-type integral over $\,{\bf p}_{\perp}(0)\,$. Performing the
subtraction at vanishing external field and going over to the variable  
$\,y\,=\,1\,-\,2\varepsilon/\omega\,$ , we find 
\be
\barl
&\dst 
T^{ij}\,=\,\frac{\alpha }{\pi}
\delta(\omega_2-\omega_1)\int\,d
{\bf r}\,exp(-i\Delbo {\bf r})\,\int_{-1}^{1}\frac{dy}{1-y^2}
\int_{0}^{\infty}\,\frac{d\tau}{\tau}\exp \left[-i\beta\tau\left(
1+\psi(\tau)\right)\right]\,B_1^{ij}\,;\\\\
&\dst
\psi(\tau)=\sum_{ {\bf q}, {\bf q}^{\prime}}\,
\frac{G({\bf q})G({\bf q}^{\prime})}{m^2}\cdot
\frac{({\bf q}_{\perp}
{\bf q}_{\perp}^{\prime})}{q_{\parallel}q_{\parallel}^{\prime}}
\,e^{i({\bf q}+{\bf q}^{\prime},\,{\bf r})}
\left[\frac{sin(q_{\parallel}+q_{\parallel}^{\prime})\tau/2}
{(q_{\parallel}+q_{\parallel}^{\prime})\tau/2}-
\frac{sin q_{\parallel}\tau/2}{q_{\parallel}\tau/2}\cdot
\frac{sin q_{\parallel}^{\prime}\tau/2}{q_{\parallel}^{\prime}
\tau/2}\right]
 \\\\ 
&\dst 
B_{1}^{ij}\,=\,y^2\,a_{1}^i\,a_{2}^j
\,-\,a_{2}^i\,a_{1}^j\,\,+
\,\delta_{\perp}^{ij}\,\left \{\,\left(\,{\bf a}_{1} {\bf a}_{2}\,\right)
\,+\,m^{2} \left[\,(y^{2}-1)\cdot\tau
\frac{d\psi}{d\tau}-\psi\,\right]\right \};\\\\
&\dst
{\bf a}_{1,2}\,=\,\sum_{\bf q}\,G({\bf q})
\cdot\frac{{\bf q}_{\perp}}{q_{\parallel}}\,e^{i{\bf q}{\bf r}}\left[\,e^{\mp i
q_{\parallel}\tau/2} -\frac{sin 
q_{\parallel}\tau/2}{q_{\parallel}\tau/2}\right]\,\,\,\,,\,\,\,\, 
\beta\,=\,\frac{2m^2}{\omega(1-y^2)}\,\,.
\ear
\aq{Trt}
\ee
This expression is still rather complicated for the numerical
calculations. It would be drastically simplified if we neglect the
function $\,\psi(\tau)\,$ as compared to $\,1\,$ in the phase of \eq{Trt}.
We suppose that the angle of incidence $\,\vartheta _0\,$ with respect 
to some major crystal axis $\,\vartheta _0\,\ll\,1\,$, since 
precisely for such angles the strengthening of electromagnetic 
effects happens in crystals as 
compared with  amorphous media. If $\,\nubo_1\,$ is far in azimuth 
from any major crystal plane, then $\,\mid q_{\parallel}\mid\, \sim 
\mid {\bf q}_{\perp}\vartheta _0\mid\,$ and   
$\,\psi_{off}(\tau)\,\leq\, (V_0/m/\vartheta _0)^2\,$. In this 
 estimate $\,V_0\,$ denotes a typical magnitude of $\,G({\bf q})\,$ 
being of the order of the axial potential well depth. So, in this 
case we can omit $\,\psi(\tau)\,$ in the phase of \eq{Trt} for 
$\,\vartheta _0\,\gg\,V_0/m\,$. If now $\,\nubo_1\,$ is aligned on 
some crystal plane, there is a subset of   $\,{\bf q}\,\,({\bf 
q}^{\prime})\,$ for which 
$\,q_{\parallel}\,\,(q_{\parallel}^{\prime})\,$ are extremely small or
vanish. Those  $\,{\bf q}\,$ are perpendicular to the plane. Expanding in 
$\,q_{\parallel}\,$ and $\,q_{\parallel}^{\prime}\,$ , we obtain for
the corresponding contribution to the phase $\,\psi_{pl}(\tau)\,=\,
(\tau /m)^2( d U_{pl}(x)/d x)^2/12\,$ . Here $\,x\,$ is 
the distance from the plane, $\,U_{pl}(x)\,$ being the planar 
potential.  As long as $\,\psi(\tau)\leq\,1\,$~, we can substitute 
$\,\tau\,\sim\, \beta^{-1}\,$ into our estimate which turns into  
$\,\psi_{pl}(\tau)\, \ll\,\kappa^2 (x)\,$. The magnitude of the 
strong field parameter $$\,\kappa (x)\,=\, 
\frac{E_{pl}(x)}{E_0}\cdot\frac{\omega}{m}\qquad,\qquad E_0\,=\,
 \frac{m^2}{e}\,=\,1.32\cdot 10^{16}\,V\,/\,cm\,$$ can be estimated using
the Table 1 of \cite {Pappla}, where, in particularly,  averaged over 
$\,x\,$ values of 
$\,(E_{pl}(x)/E_0)^2\,$ are presented for $\,(110)\,$ plane of several
crystals. As a result, for commonly used crystals, $\,\psi_{pl}(\tau)\,$ is 
sufficiently small when $\,\omega\,$ is less than several $\,TeV\,$.
Assuming that $\,\omega\,$ and $\,\vartheta _0\,$ satisfy the formulated
conditions, we, finally, rule out $\,\psi(\tau)\,$ from the phase of \eq{Trt} .

Now the integration over {\bf r} in \eq{Trt} can be performed 
$$\int\,d
{\bf r}\,exp \left[i({\bf q}\,+\,{\bf q}^{\prime}-\Delbo\,\,,\,\, {\bf r})
\right]\,
=\,(2\pi)^3\,\delta({\bf q}\,+\,{\bf q}^{\prime}-\Delbo)\,.$$ As a result,
the tensor $\,T^{ij}\,$ acquires the form
$$ T^{ij}(k_{1},k_{2})\,=\,(2\pi)^{3}\,
\sum_{{\bf Q}}\pi^{ij}({\bf Q}\,,k_{1})\,\delta (k_{2}\,-\,k_{1}\,-\,Q),$$
where $\,Q \,=\,{\bf q}\,+\,{\bf q}^{\prime}\,$. Using this presentation of
$\, T^{ij}(k_{1},k_{2})\,$  and
\eq{Fcr},\eq{Hit} we find that $\,c^{ij}({\bf Q}\,,k_{1})\,$ in \eq{Fcr}
are $\,c^{ij}({\bf Q}\,,k_{1})\,=\,\pi^{ij}({\bf Q}\,,k_{1})/(4\pi\omega^2)
\,$. Here we are interested only in $\,\pi^{ij}(0\,,k_{1})\,$ being just
the polarization operator $\,\Pi\,$ which 
determines a development of the photon density matrix at large thicknesses
according to \eq{Rho}  .
Taking elementary integrals over $\,\tau\,$ and $\,y\,$, we obtain from
\eq{Trt}
\be
\Pi^{ij}=\frac{\alpha \omega^2}{8m^2}
\sum_{ {\bf q}}\frac{\mid G({\bf q})\mid ^2}{m^2}
\left \{{\bf q}_{\perp}^2\delta_{\perp}
^{ij}
\left[id_1(\mu)+d_2(\mu)\right]+\left(\frac{1}{2}{\bf q}_{\perp}^2
\delta_{\perp}^{ij}
-q_{\perp}^i q_{\perp}^j 
 \right)\left[id_3(\mu)-d_4(\mu)\right]\right \}\,, 
\aq{Tfin} 
\ee 
where $\,\mu\,=\,2 m^2/(\omega \mid 
q_{\parallel}\mid)\,$ and functions $\,d_l\,$ are 
\be 
\barl 
&\dst 
d_1(x)\,=\,x^2\left[\,\left(1+x-\frac{1}{2}x^2\,\right)g(\sqrt {1-x})-(1+x)
\sqrt {1-x}\,\right] \,\theta (1-x)\quad , \\ \\
&\dst 
 d_3(x)\,=\,x^3\left(\,\frac{x}{2}g(\sqrt {1-x})+\sqrt {1-x}\,\right)
  \,\theta (1-x)\quad , \qquad g(x)=\ln \frac {1+x}{\mid 1-x \mid }\quad ,\\ \\
&\dst
d_4(x)\,=\,\frac{x^2}{\pi}\Big\{ \left( \,\frac{x}{2}
 g(\sqrt{1+x})-\sqrt{1+x}\, \right)^2
- \left(\,x \arctan \frac{1}{\sqrt {x-1}}-\sqrt {x-1}\,\right)^2 \theta (x-1)+ 
\\ \\
&\dst
+\left[\left(\,\frac{x}{2}
 g(\sqrt {1-x})+\sqrt {1-x}\,\right)^2-\left(\,
 \frac{\pi x}{2}\,\right)^2 \right]\,\theta (1-x) \Big\}
\,\,,\\ \\
&\dst
d_2(x)\,=\,d_4(x)+ \frac{2x^2}{\pi}\Big\{ \frac{1}{2}g(\sqrt{1+x})\,
\left(\, \frac{x-1}{2}\,g(\sqrt{1+x})+\sqrt{1+x}\,\right)-3-
\\\\
&\dst
-\left[\,\frac{1}{2}g(\sqrt{1-x})\,
\left(\, \frac{x+1}{2}\,g(\sqrt{1+x})-\sqrt{1-x}\,\right)
+1-\frac{\pi^2(1+ x)}{4}\,\right]\,\theta (1-x)+
\\\\
&\dst
+\left[\,\arctan \frac{1}{\sqrt {x-1}} \left( \,(x+1)\arctan \frac{1}{\sqrt {x-1}}
+\sqrt {x-1}\,\right)-1\right]\,\theta (x-1)\Big\}\,\,,
\ear
\aq{Did}
\ee
where $\,\theta (x)\,=\,1\,$ for $\,x\,>\,0\,$ and vanishes for $\,x\,<\,0\,$.

The photon energy $\,\omega\,$ enters \eq {Tfin} in the combination 
$\,\mu\,$ except of the factor $\,\omega^2\,$ in front of the sum. 
 This sum is almost independent of $\,\omega\,$ near its maximum 
since the latter is given by $\,\mu\,\simeq\,1\,$. As 
a result, the $\,\omega\,$-dependence of the quantities $\,\alpha\,$ 
and $\,\beta\,$ in \eq {Rof} and \eq {expl} describing the behavior 
of the photon polarization is reduced to the factor $\,\omega L\,$ 
for optimal orientations. Correspondingly, the optimal thickness
$\,L_{opt}\,$ is roughly proportional to $\,\omega^{-1} \,$.
 As mentioned above, there is a subset of 
$\,{\bf q}\,$ at perfect planar alignment for which 
$\,q_{\parallel}\,$ vanishes ($\,\mu \rightarrow \infty\, $).  The 
contribution of this subset to $\,\Pi^{ij}\,$ , i.e.  the yield of 
the corresponding planar potential obtained from \eq {Tfin} 
reproduces well known results ( see, e.g.  \cite {Pappla} and 
literature cited there ) derived within the Born approximation:  
$$ 
\frac {\alpha m^2}{45 \pi}\,\langle \kappa^2 (x)\,\rangle\,\Big( 
4\,e_1^ie_1^j\,+\,7\,e_2^ie_2^j\,\Big) \qquad, 
$$ where $\,\langle 
\dots\rangle\,$ means averaging over the coordinate $\,x\,$ , $\,{\bf 
e}_1\,$ is the unit vector perpendicular to the plane and $\,{\bf 
e}_2\,=\,\nubo_1\times {\bf e}_1\,$ . What is lost when  
$\,\psi(\tau)\,$  has been omitted in the phase of \eq{Trt}, are 
higher order corrections in crystal potential .

Using the explicit form of $\,\Pi^{ij}\,$ \eq {Tfin} and adding the 
incoherent yield,we can find the quantities presented in \eq{Rof}, which 
describes the properties of a
photon beam for any initial conditions dependent on the crystal thickness.
As explained above, the incoherent yield is present in \eq{Rof} only in the
absorption coefficient $\,\gamma\,$. We can use any basis to calculate
$\,\Pi^{ij}\,$. Let this basis be formed by two real unit vectors $\,{\bf e}_1
\,$ and $\,{\bf e}_2\,$ satisfying $\,{\bf e}_1{\bf e}_2\,=\,\nubo_1{\bf e}_1
\,=\,\nubo_1{\bf e}_2\,=\,0\,$. We should use the same basis to obtain the
initial Stokes vector $\,\ebol\,$. Supposing that $\,\nubo_1\,$ is near some axis
direction $\,\nubo_3\,$, we choose $\,{\bf e}_2\,$ in the plane ( reaction plane )
containing $\,\nubo_1\,$ and $\,\nubo_3\,$. For the basis chosen,
the circular polarization degree is given by the magnitude of the second Stokes
parameter $\,\xi(L)\,=\,\mid \eta_2 (L)\mid\,$.
If we define the angle of incidence $\,\vartheta_0\,$ as the angle 
between $\,\nubo_1\,$ and $\,\nubo_3\,$, and $\,\varphi_0\,$ which is 
the angle between the reaction plane and the $\,(1 \bar {1}0)\,$ 
plane, then the explicit form of the basis vectors reads 
$$ {\bf e}_1\,=\,-{\bf e}_x \sin \varphi_0\,+\,{\bf e}_y \cos 
\varphi_0 \quad,\quad {\bf e}_2\,=\,\nubo_3 \sin \vartheta_0\,-\, 
({\bf e}_x \cos \varphi_0\,+\,{\bf e}_y \sin \varphi_0)\cos 
\vartheta_0\,,$$ 
where $\,{\bf e}_x{\bf e}_y\,=\,\nubo_3{\bf e}_x
\,=\,\nubo_3{\bf e}_y\,=\,0\,$ and $\,{\bf e}_x\,$ is in the
 $\,(1 \bar {1}0)\,$  plane. In this basis the quantities
$\,a\,$ and $\,{\bf b}\,$  presented in \eq{Rof}  take the form
\be
\barl 
&\dst 
( a, {\bf b} )\,=\,\frac{\alpha \omega}{16m^2}
\sum_{ {\bf q}}\frac{\mid G({\bf q})\mid ^2}{m^2}
( A,{\bf B} )\quad,\\\\
&\dst 
A\,=\,{\bf q}_{\perp}^2 \left[id_1(\mu)+d_2(\mu)\right]\quad,\quad
B_1\,=\,-({\bf e}_1 {\bf q}_{\perp})({\bf e}_2 {\bf q}_{\perp})
\left[id_3(\mu)-d_4(\mu)\right]\,,\\\\
&\dst 
B_2\,=\,0\quad, \quad
B_3\,=\,\frac{1}{2}\left[({\bf e}_2 {\bf q}_{\perp})^2\,-\,({\bf e}_1 
{\bf q}_{\perp})^2\right] \left[id_3(\mu)-d_4(\mu)\right]\,.
\ear 
\aq{abex} 
\ee 
Remember that the eigenvalues of the propagation matrix
$\,\Pi/(2 \omega )\,$ in \eq {Rho} are  $\,a\,\pm\,\sqrt {{\bf b}^2}\,$.
The corresponding eigenvectors $\,{\bf e}_{\pm}\,$ satisfying the
normalization condition $\,\mid {\bf e}_{\pm} \mid^2\,=\,1\,$ are
\be
{\bf e}_{+}\,=\,\frac{{\bf e}_1\,+\,r{\bf e}_2}{\sqrt {1+\mid r \mid 
^2}}\quad,\quad 
{\bf e}_{-}\,=\,\frac{{\bf e}_2\,-\,r{\bf e}_2}{\sqrt {1+\mid r \mid 
^2}}\quad,\quad r\,=\,\frac{b_1}{b_3+\sqrt {{\bf b}^2}}\,.
\aq{eigen} 
\ee 
In general, the  eigenvectors $\,{\bf e}_{\pm}\,$  are complex. 
However, they become real when the quantity $\,r\,$ does so. In 
 particularly, when $\,\nubo_1\,$ is in the symmetry plane of a 
 crystal like $\,(1 \bar {1}0)\,$ plane ($\,\varphi_0\, =\,0\,$),
$\,r\,$ vanishes and $\,{\bf e}_{\pm}\,$  coincide with
 $\,{\bf e}_{1,2}\,$. Just such a case  ($\,\varphi_0\, 
 =\,\pi/2\,$,$\,\nubo_3\,$ along $<110>$ axis) was the only 
orientation considered
in \cite {Cabibo} where a quantity accounting for 
the polarization conversion was calculated by the use of dispersion 
relations. In our notation it corresponds to the term in \eq{Tfin} 
proportional to $\,d_4(\mu)\,$ . We were unable to reproduce the factor 
in eq.(5) of \cite {Cabibo}, nevertheless, we emphasize that
the expression in braces of the cited equation coincides with  
$\,d_4(\mu)/\mu^2\,$.
 
 Consider now, as an example, fully linearly polarized ($\, \eta_2 
\,=\,0\, , \, \ebol^2\,=\,1\,$) initial photon beam. An efficiency of 
the polarization conversion process is determined not only by 
$\,\xi(L)\,$ but also by the fraction of surviving photons 
$\,FRAC(L)\,$.  We use the criterion of such an  efficiency suggested 
in \cite {Proposal} $$FOM(L)\,=\,10 \cdot\xi(L) \sqrt {FRAC(L)}\quad 
,$$ $\,L\,$ being the crystal thickness. Recollect that $\,\xi(L)\,$ 
 and $\,FRAC(L)\,$ depend also on $\,\omega\,$ and $\,\nubo_1\,$.
For given orientation of a crystal ,  we 
still have a free parameter which is the angle $\,\phi\,$ of the initial 
polarization vector with respect to the basis vectors $\,{\bf e}_1\,$ and 
 $\,{\bf e}_2\,$. We do not
claim here to the final analysis of the polarization conversion process, 
so that
the yield for different orientations will be compared at the same initial
condition. Namely, we set $\,\eta_1 \,=\,-1\,$ which corresponds to
$\,\phi\,=\,\pi/4\,$ with respect to $\,{\bf e}_2\,$
( $\,\phi\,=\,3\pi/4\,$ with respect to $\,{\bf e}_1\,$).
 Evidently, this
is the best choice for the alignment on the symmetry plane of a 
 crystal  when the basis vectors are 
 the eigenvectors of the matrix $\,\Pi^{ij}\,$ as well. 

In Fig.1 the maximum values of the figure of merit ( $\,FOM\,$ )  
are shown near $<110>$ axis of a diamond crystal
for $\, \omega\, =\,100\,  GeV\,$ as a function of the orientation 
(angles $\,\vartheta_0\,$ and $\,\varphi_0\,$) .  For each 
direction $(\,\vartheta_0\,, \,\varphi_0\,)$, the 
figure of merit was calculated first as a function of $\,L\,$ , then 
its maximum value was found. Just these maximum values of FOM are 
plotted, so that different directions correspond 
usually to different optimal thicknesses $\,L_{opt}\,$.  Owing to the 
crystal symmetry, there is no need to perform calculations for 
$\,\varphi_0\,$ beyond the interval chosen, since they will simply reproduce
the results already obtained for corresponding $\,\varphi_0\,$ within the
interval. For diamond crystal, the largest effect is achieved for 
$\,\vartheta_0\, =\,1.5\,mrad\,$ off the
$\,<110>\,$ axis on the $\,(1\bar{1}0\,$ plane ($\,\varphi_0\, =\,0\,$), where
 ( see Fig.1 ) $\,FOM\,\simeq\,3.7\,$. However, in this case the size 
of $\,L_{opt}\,\simeq\, 4.7\,cm\,$ seems to be too large for 
 practical use. Generally, a typical size of the optimal thickness is 
of a few $\,cm\,$. For the same axis of a silicon crystal ( see Fig.2 
) , the peak is at $\,\vartheta_0\, =\,2.3\,mrad \,$ and 
$\,\varphi_0\, =\,0\,$ with $\,FOM\,\simeq\,2.6\,$. We emphasize 
rather narrow angular with of the peak in both directions. The 
optimal thickness in this case is $\,L_{opt}\,\simeq\,8\,cm\,$.  We 
have performed the same kind of calculations for three major axes of 
diamond, silicon and germanium crystals. Comparing types of crystals, 
we see that the effect is the largest for diamond and the smallest 
for germanium, where $\,FOM\,\simeq\,2.0\,$ can be obtained at 
$\,L_{opt}\,\simeq \,2.8\,cm\,$. What about orientations, the 
$\,<110>\,$ axis is the most preferable, while two others give a 
comparable but smaller yield.

For $\,\varphi_0\, =\,0\,$  the position of a 
peak is determined by the condition $\,\mu\,=\,2 m^2/(\omega \mid 
q_{\parallel}\mid~)\,=\,1\,$ for the smallest non-zero $\,\mid 
q_{\parallel}\mid\,$. This is connected with the threshold 
behavior of the functions $\,d_{1,3}(\mu)\,$ in \eq {Tfin} at 
$\,\mu\,=\,1\,$.  From this condition we obtain for $\,fcc(d)\,$ 
structure near $\,<110>\,$ axis $\,\vartheta_0^{max}\, =\,m^2l_c/(\pi 
\omega) \,$ , where $\,l_c\,$ is a lattice constant. In particularly, 
for $\,Si\,$ we have $\,\vartheta_0^{max}(mrad)\, \simeq\,229/ 
\omega(GeV) \,$.  This fact is illustrated by Fig.3 where $\,FOM\,$ 
(upper curves), $\,FRAC\,$ (lower curves), and the degree of circular 
polarization $\,\xi\,$ are shown for $\,\varphi_0\, =\,0\,$ near 
$\,<110>\,$ axis of a $\,10 cm\,$ thick silicon crystal as functions 
of $\,\omega\,$.  Three sets of curves ( from the left to the right 
) in Fig.3 correspond to the angles of incidence 
$\,\vartheta_0(mrad)\, =\,$ 2.29 , 2.08 and 1.91 respectively. All  
curves in Fig.3 have peaks exactly at the positions prescribed by the 
condition obtained above.  For a given orientation ( at fixed 
$\,\vartheta_0\,$ ), rather narrow shape of these peaks does not 
allow us to handle with the same efficiency a photon beam having the 
wide energy spread.

The magnitude of $\,FOM(L)\,$ being proportional to  $\,\xi(L)\,$  
diminishes for partially polarized initial photon beam roughly
proportionally to $\,\mid \ebol 
\mid\,<\,1\,$  as compared to the fully polarized  case ($\,\mid \ebol 
\mid\,=\,1\,$ ). Depending on the orientation, more or less noticeable
change of the polarization degree $\,\mid \ebol(L) \mid\,$ occurs for
$\,\mid \ebol \mid\,<\,1\,$. It can be seen in Figs.4,5 where the
absolute values of three Stokes parameters and the polarization degree
are presented as functions of the silicon crystal thickness $\,L\,$
at $\,\omega\,=\,100 GeV\,$.
The calculations were carried out using \eq {etaf}, \eq {expl}, and
 \eq {abex} for $\,\eta_1 \,=\,-0.5\,$ , $\,\eta_2 \,=\,\eta_3\,= 
\,0\,$. The angle of incidence $\,\vartheta_0\, =\,2.29\,mrad \,$ 
is the same for both figures. At $\,\varphi_0\, =\,0\,$ ( Fig.4 ),
$\,\eta_3(L)\,$ is small and $\,\mid \ebol 
(L)\mid\,$ ( curve "tot" ) practically does not change due to the
smallness of the parameter $\,\alpha\,$ ( see \eq {Rof}) within the
whole interval of $\,L\,$ presented. The fracture of the curve (1) 
means only that $\,\eta_1(L)\,$ changes its sign at $\,L\,=\,21 
cm\,$. Since the polarization degree practically conserves and 
$\,\eta_3(L)\,$ can be neglected for this orientation, we could
measure the linear polarization $\,\eta_1(L)\,$ to determine the
circular  polarization $\,\eta_2(L)\,$ appeared. However,
the situation  drastically changes already for a small alteration of
$\,\varphi_0\,$. It is seen in Fig.5 calculated at $\,\varphi_0\, =
\,0.015\,$. Now the measurement of the linear polarization does not
help in the determination of the circular one since 
the polarization degree  is no more constant. 

In conclusion, formulas derived describe the propagation 
of hard polarized photons through crystals. They are valid in
a wide photon energy range for any orientation and any crystal type 
as long as the approximations used are correct. Our calculations
show that the linear polarization of multi-GeV photons can be
converted with an appropriate efficiency  into the circular one using 
properly chosen single crystals. However, if we do not use
theoretical results for some quantities involved (~e.g. for
the polarization degree~), the only way to determine the
circular  polarization appeared in a crystal is the direct measurement of
it.

\vspace{1.0cm}
{\bf Acknowledgements} 

\vspace{0.5cm}
The author is grateful to V.M.Katkov and A.I.Milstein for 
many fruitful discussions.

\newpage
\begin{center}
{\bf Figure captions}
\vspace{15mm}
\end{center}
\begin{enumerate}

\item {\bf Fig.1} The maximum values of  $\,FOM\,$ ( the figure of merit
, see text ) near $\,<110>\,$ axis of a diamond crystal for $\,
\omega\, =\,100\,  GeV\,$ dependent on the angles of incidence $\,
\vartheta _0\,$ and $\,\varphi _0 \,$.

\item {\bf Fig.2} The same as in Fig.1 but for a silicon crystal
in the neighborhood of the point $\,\vartheta _0\,=\,2.3 mrad\,$ , 
$\,\varphi _0 \,=\,0\,$.

\item {\bf Fig.3} $\,FOM\,$ 
(upper curves), $\,FRAC\,$ (lower curves), and the degree of circular 
polarization $\,\xi\,$  for $\,\varphi_0\, =\,0\,$ near 
$\,<110>\,$ axis of a $\,10 cm\,$ thick silicon crystal as functions 
of $\,\omega\,(GeV)\,$. Three sets of curves ( from the left to the 
right ) correspond to the angles of incidence $\,\vartheta_0(mrad)\, 
=\,$ 2.29 , 2.08 and 1.91 respectively.

\item {\bf Fig.4}  Absolute values of the Stokes parameters 
$\,\mid \eta_1 (L)\mid\,$ (curve 1),
$\,\mid \eta_2 (L)\mid\,$ (curve 2),
$\,\mid \eta_3 (L)\mid\,$ (curve 3), and 
the polarization degree $\,\mid \ebol (L)\mid\,$ (curve "tot"~)
depending on the  thickness 
$\,L\,$ of a silicon crystal at $\,\omega\,=\,100 GeV\,$,
$\,\eta_1 \,=\,-0.5\,$ , $\,\eta_2 \,=\,\eta_3\,= 
\,0\,$, $\,\vartheta_0\, =\,2.29\,mrad \,$, 
$\,\varphi_0\, =\,0\,$ .

\item {\bf Fig.5} The same as in Fig.4 but for $\,\varphi _0 
\,=\,0.015\,$.

\end{enumerate}
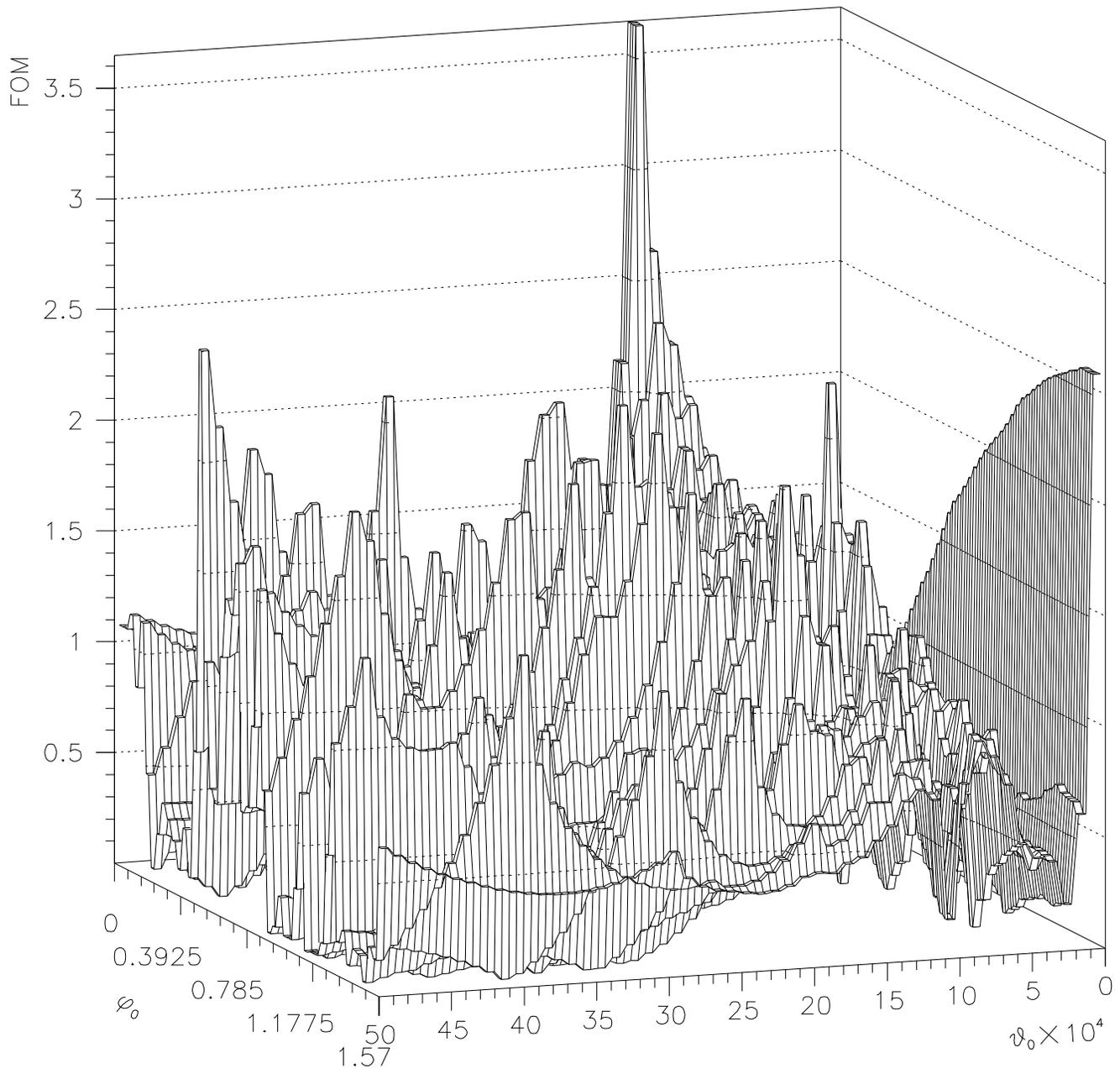
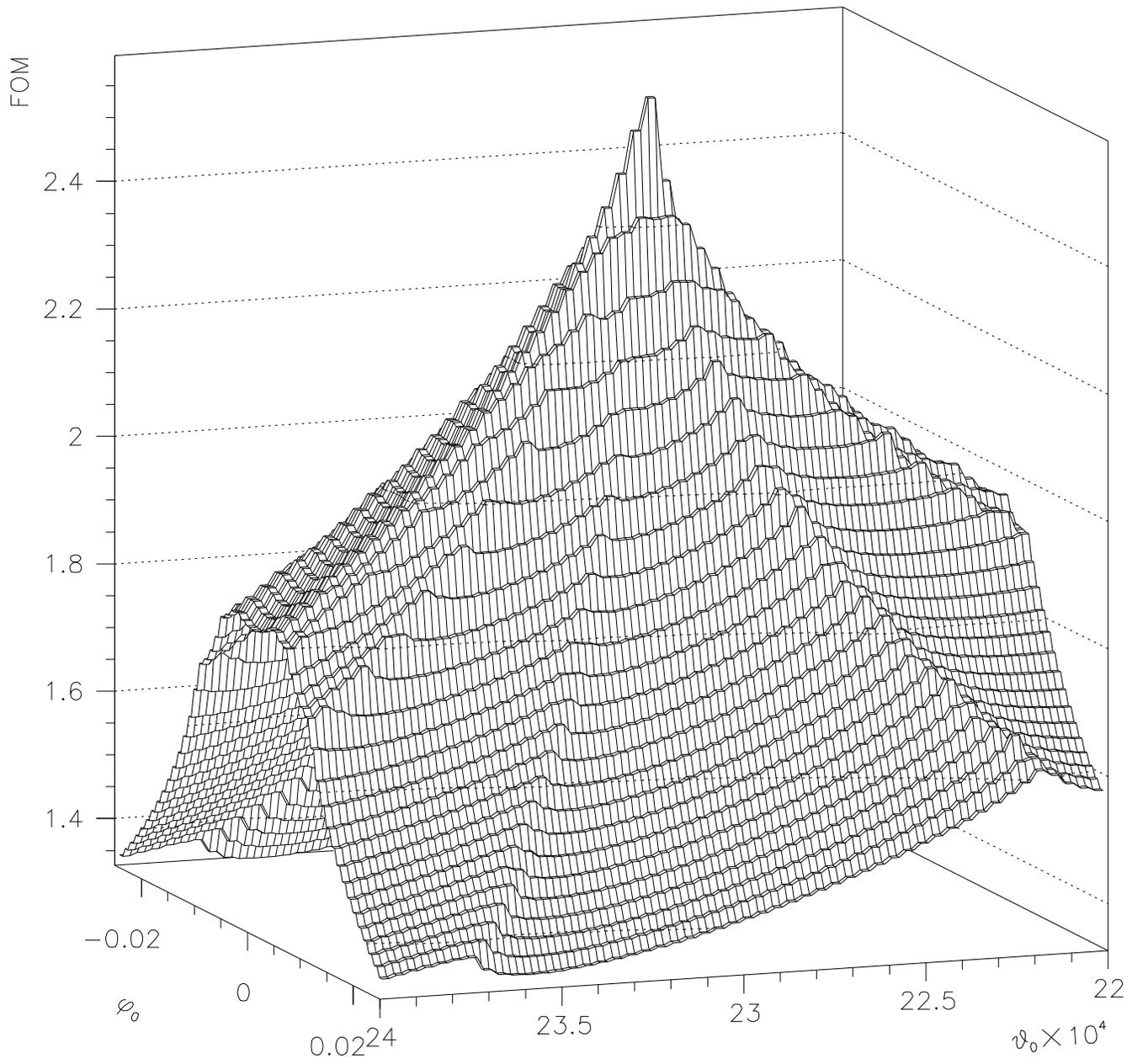
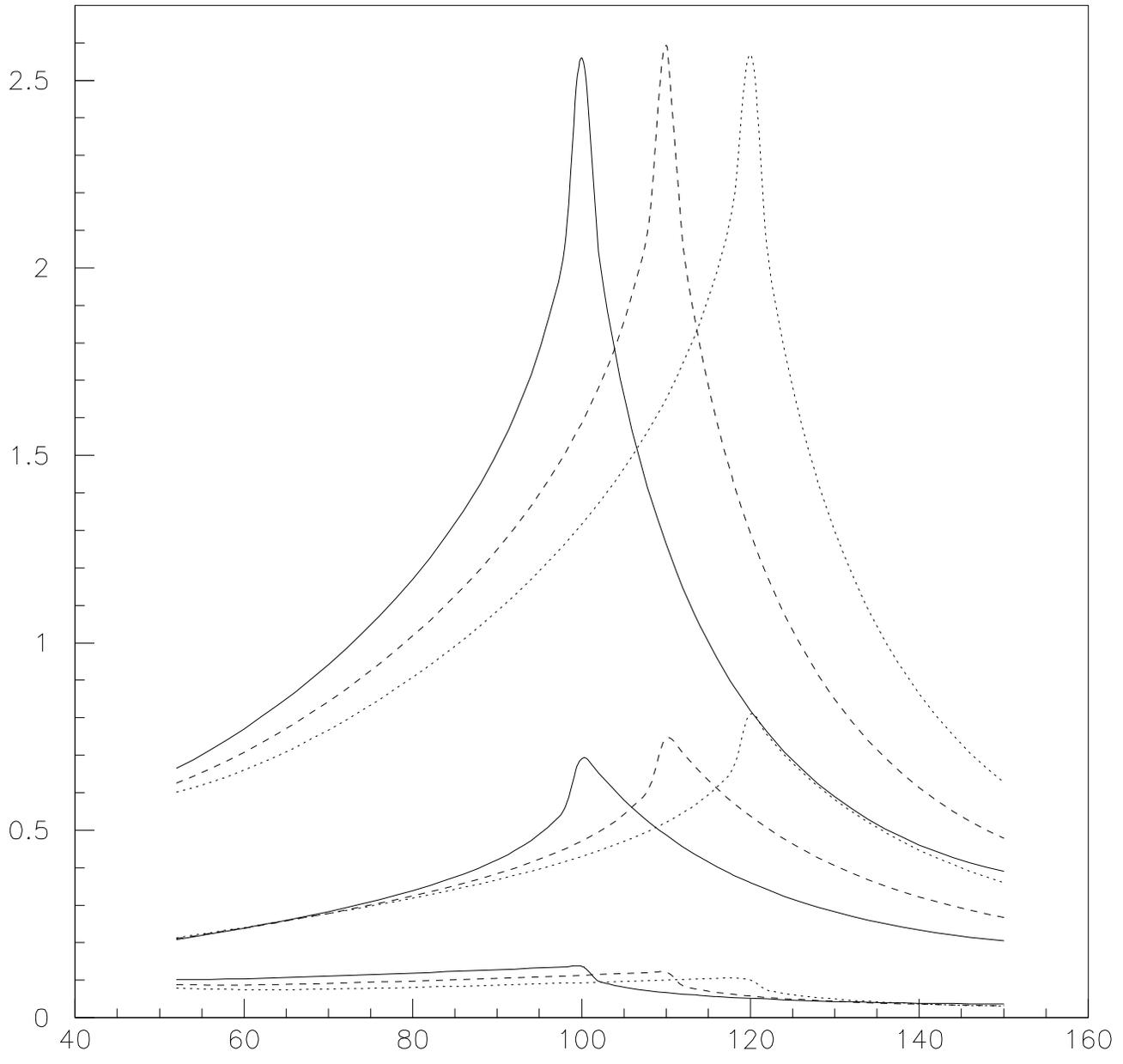
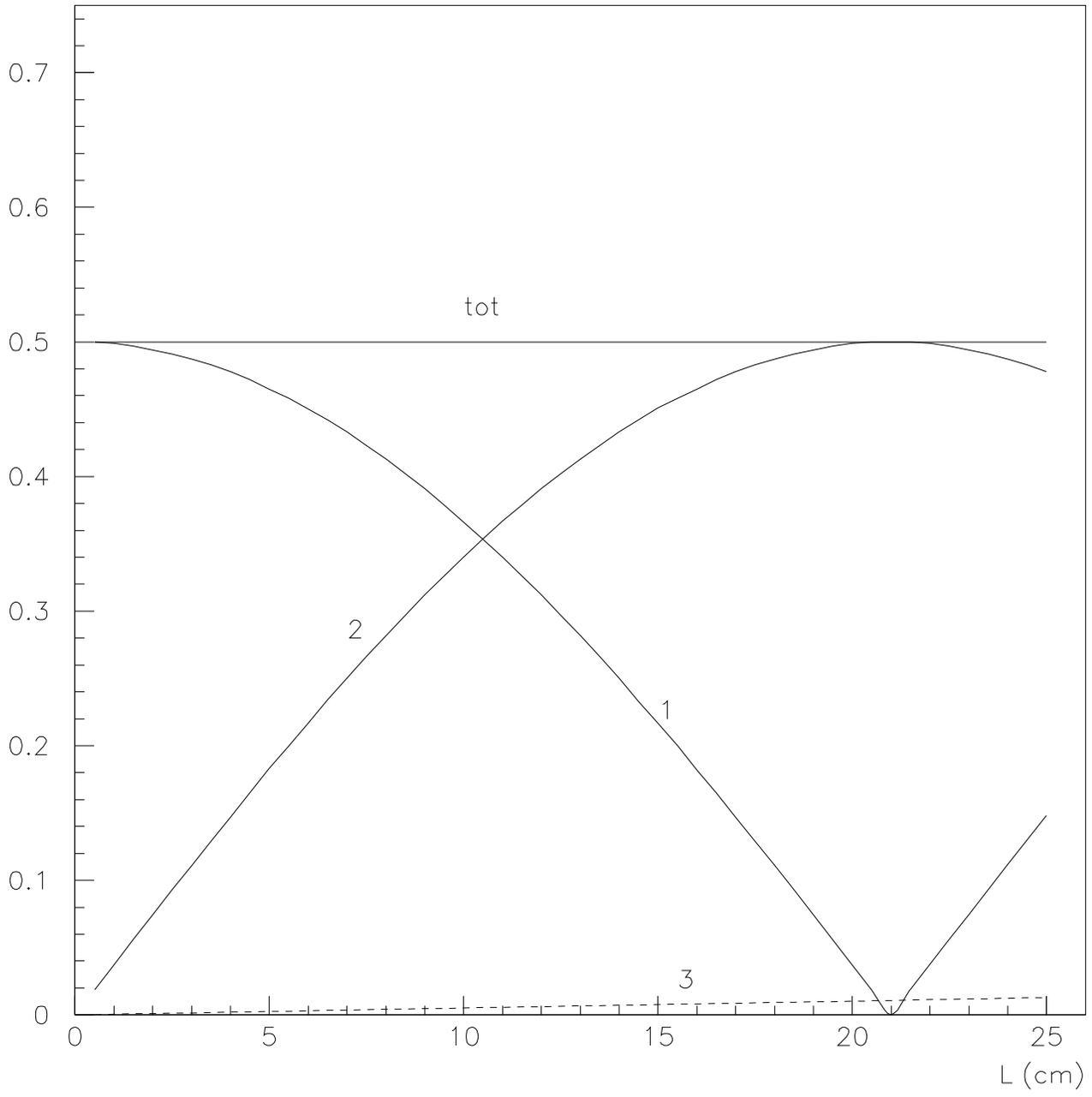
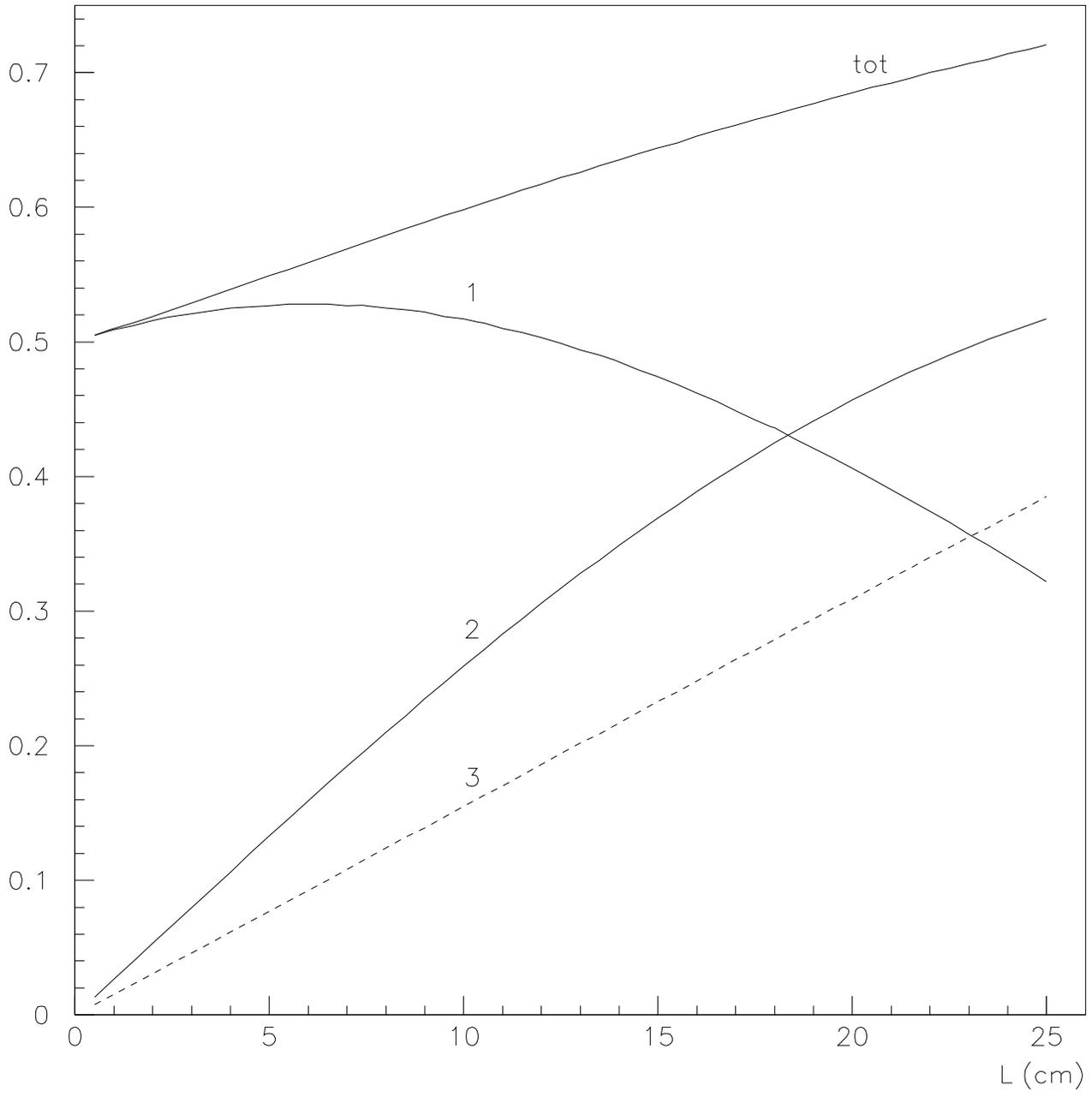
\end{document}